\documentstyle[12pt]{article}

\newlength{\numlen}

\newlength{\indexlength}

\newcommand{\eq}{\begin{equation}}
\newcommand{\en}{\end{equation}}

\newcommand{\be}{\begin{equation}}
\newcommand{\ee}{\end{equation}}
\newcommand{\ba}{\begin{eqnarray}}
\newcommand{\ea}{\end{eqnarray}}

\begin{document}

\begin{titlepage}

\hfill CERN-TH.7080/93\\

\begin{centering}
\vfill
{\large \bf Strong Sphalerons and Electroweak Baryogenesis}\\
\vspace{1cm}
G.~F. Giudice\footnote{On leave of absence from INFN, Sezione di
Padova.}
and M. Shaposhnikov\footnote{On leave of
absence from the Institute for
Nuclear Research of the Russian Academy of Sciences, Moscow 117312,
Russia.} \\
\vspace{0.5cm}
{\em Theory Division, CERN,\\ CH-1211 Geneva 23, Switzerland}\\
\vspace{0.3cm}
{\bf Abstract}\\
\vspace{1cm}
\end{centering}
We analyze the spontaneous baryogenesis and
charge transport mechanisms suggested by Cohen, Kaplan and Nelson for
baryon asymmetry generation in extended versions of electroweak
theory. We find that accounting for non-perturbative chirality-breaking
transitions due to strong sphalerons reduces the
baryonic asymmetry by the factor $(m_t/\pi T)^2$ or $\alpha_W$,
provided those processes are in thermal equilibrium.
\vspace{0.3cm}\noindent

\vfill \vfill
\noindent CERN-TH.7080/93\\
\noindent November 1993
\end{titlepage}

The popular mechanisms for electroweak baryogenesis
discussed in the literature are the spontaneous baryogenesis
mechanism and the charge transport mechanism, both suggested
by Cohen, Kaplan and Nelson \cite{ckn1,ckn2,ckn3}. It has been
advocated in
\cite{ckn1,ckn2,ckn3} that these mechanisms produce an asymmetry
parametrically larger that the mechanism associated with the decay
of topologically non-trivial fluctuations of the gauge and Higgs
fields during the electroweak phase transition, considered in
refs. \cite{mstv, gst}(for earlier suggestions, see refs.
\cite{ms,amb,neil}).
In the latter
case, the asymmetry produced is proportional to the square of the ratio
of the top-quark mass to the temperature \cite{mstv}, while for the
first
two
mechanisms it was argued that this suppression is absent
\cite{ckn1,ckn2,ckn3}. The key observation made in
\cite{ckn1,ckn2,ckn3} is that in the presence of a non-zero fermionic
hypercharge density, $\rho_Y$, the equilibrium value of the baryonic
charge, generally speaking, is not equal to zero:
\eq
\langle B \rangle = A \rho_Y,
\label{bar}
\en
where $A$ is a number determined by the particle content of the theory
and $A \neq 0$ in the limit $m_t \rightarrow 0$.

In this paper, we show that under some circumstances $A$ may be zero in
the massless quark limit. Moreover, we will argue that this is most
probably the case for the two Higgs doublet theory, considered in
\cite{ckn1,ckn2}.

In this model, in addition to the KM CP-violation, there is an explicit
CP-violation in the Higgs sector. The vevs of the Higgs fields at zero
temperature have the form:
\eq
\langle \phi_1 \rangle = (0, v_1 e^{i \theta}),~
\langle \phi_2 \rangle = (0, v_2),
\en
where $\theta$ is the relative phase, which cannot be rotated away by
a gauge transformation.
Due to explicit CP-violation in the Higgs potential this phase is
space-dependent inside bubble walls forming at the electroweak phase
transition (it would be time-dependent for a spinodial decomposition
phase transition). Since quarks are getting their masses from
interactions with
the Higgs fields, the most important interaction  is the top-quark
coupling\footnote{We assume that the first Higgs field couples
only to up-quarks, while the second is coupled only to down-quarks.
This assumption is, however, inessential in what follows.},
\eq
L_Y = f_t \bar{Q_3}U_3 \phi_1.
\label{yukawa}
\en
Here $Q_i$ are the left fermionic doublets, $U_i,~D_i$ are right
quark fields, and $i$ is the generation index.
The bottom-quark couplings can be safely ignored provided the thickness
of
the domain wall $l_0$ or the duration of the spinodial decomposition
phase
transition $\tau_0$ is smaller than the rate of chirality-changing
transitions associated with the bottom Yukawa coupling; that is, if
\eq
\l_0 < (\alpha_s f_b^2 T)^{-1} \sim \frac{10^4}{T},
\en
where $T$ is the temperature. This is always the case for current
estimates of the domain wall thickness, ranging from $40/T$
\cite{dine} to $1/T$ \cite{ms:wall}. The same statement is obviously
true for lighter quarks.

We will clarify our point first within the context of the spontaneous
baryogenesis mechanism.
Let us take for simplicity the spinodial decomposition phase
transition. As in \cite{ckn1}, we assume that the rate of
chirality-flipping
transitions for the top quark is larger than the typical inverse-time
scale $\tau_0^{-1}$ of the
variation of the phase $\theta$ during this transition. Now, following
\cite{ckn1}, we make a hypercharge rotation of the fermionic fields in
such
a way that time-dependence disappears from the Yukawa coupling
(\ref{yukawa}). This change of variables introduces the following
modification of the effective lagrangian:
\eq
L_{eff} = L + \dot{\theta}Y_F,
\label{eff}
\en
where $Y_F$ is a fermionic hypercharge operator,
\eq
Y_F = \sum_{i=1}^3\left[ \frac{1}{3}\bar{Q_i}\gamma_{0}Q_i
+ \frac{4}{3}\bar{U_i}\gamma_{0}U_i
- \frac{2}{3}\bar{D_i}\gamma_{0}D_i
- \bar{L_i}\gamma_{0}L_i
- 2 \bar{E_i}\gamma_{0}E_i\right] .
\en

We want to check whether in thermal
equilibrium with respect to fermionic number non-conserving processes
in a theory with effective action (\ref{eff}) the baryonic number is
nonzero. To determine the
number $A$ defined in (\ref{bar}), the following standard procedure
must be
used (see {\em e.g.} ref. \cite{landau}). One
has to define the {\em complete} set of conserved charges $X_i$ and
construct the most general equilibrium density matrix with the help of
those charges, introducing chemical potentials $\mu_i$ for each of
them. The partition function is
\eq
Z  = Tr\exp{\left[ - \frac{1}{T}(H -  \dot{\theta}Y_F - \sum_i \mu_i
X_i)
\right] },
\label{sum}
\en
while the density matrix is
\eq
\rho  = \frac{1}{Z}\exp{\left[ - \frac{1}{T}(H -  \dot{\theta}Y_F -
\mu_i
X_i)\right] }.
\label{density}
\en
Now, since $X_i$ are conserved operators, their average must be
equal to zero. This requirement fixes the chemical potentials $\mu_i$
through the system of equations
\eq
\frac{\partial}{\mu_i}Z = 0.
\label{equations}
\en
Then the baryonic number is just
\eq
\langle B \rangle = Tr[B \rho].
\label{baryon}
\en

Let us define the set of conserved numbers. It is sufficient to
consider only SU(3) gauge singlets. We start from purely
fermionic currents. In a model with massless neutrinos (no right-handed
neutrinos!), the total number of different fermionic currents is 15,
the number of fermionic degrees of freedom. They are: 3 left quark
currents, $\bar{Q_i}\gamma_{\mu}Q_i$, 6 right quark currents,
$\bar{U_i}\gamma_{\mu}U_i$ and  $\bar{D_i}\gamma_{\mu}D_i$, 3 left
leptonic currents, $\bar{L_i}\gamma_{\mu}L_i$ ($L_i$ is the left
leptonic doublet), and 3 right leptonic currents,
$\bar{E_i}\gamma_{\mu}E_i$. Not all of these currents are conserved.
One
has to take into account the following processes:\\

(i) Perturbative chirality-changing transitions due to Yukawa
interactions
(\ref{yukawa}). The rate of these processes is estimated to be $\tau_Y
\sim 30/T$ \cite{ckn3}. This decreases the number of conserved currents
by
one: instead of the pair  $\bar{Q_3}\gamma_{\mu}Q_3$ and
$\bar{U_3}\gamma_{\mu}U_3$, we have a conserved linear combination
$\bar{Q_3}\gamma_{\mu}Q_3 + \bar{U_3}\gamma_{\mu}U_3$.  \\

(ii) Non-perturbative chirality-breaking transitions due to strong
interactions. It is well known that the quark axial vector current has
an
anomaly and therefore is not conserved. The rate of chirality
non-conservation at high temperatures $\Gamma_{strong}$ is connected
with the rate of topological transitions in QCD (``strong''
sphalerons, \cite{mms}),
\eq
\frac{\partial Q_5}{\partial t} = - \frac{12\cdot
6}{T^3}\Gamma_{strong}Q_5,
\en
where $Q_5$ is the axial charge. The factor of $12$ comes from the
total number of quark chirality states, the factor of $6$ from the
relation between
the asymmetry in quark number density and the chemical potential,
\eq
\int\frac{d^3 k}{(2\pi)^3}[n_F(\epsilon,\mu) - n_F(\epsilon,-\mu)]=
\frac{\mu T^2}{6}(1-\frac{3 m^2}{2 \pi^2 T^2}),
\label{quark}
\en
where $n_F(\epsilon,\mu) = 1/(\exp (\frac{\epsilon - \mu}{T}) +1)$ is
the Fermi distribution, $\epsilon^2 = k^2 + m^2$. The rate of the
strong sphaleron transitions is related to the rate of weak sphaleron
transitions through
\eq
\Gamma_{strong} = \frac{8}{3}(\frac{\alpha_s}{\alpha_W})^4
\Gamma_{sph} = \frac{8}{3}\kappa(\alpha_s T)^4,
\en
where $\kappa$ is the usual parameter characterizing the strength of
the electroweak sphaleron transitions in the unbroken phase. The
characteristic time of these transitions is therefore
\eq
\tau_{strong} = \frac{1}{192 \kappa \alpha_s^4 T}.
\en
Using the conservative lower bound on $\kappa$ derived in lattice
simulations \cite{aaps}, $\kappa > 0.5$ (see also discussion in ref.
\cite{far}), we obtain a conservative bound $\tau_{strong}< 100/T$.
With
$\kappa \sim 20$, so estimated by a different method in ref.
\cite{armc},
we obtain
$\tau_{strong} \sim 2.5/T$. Therefore, the rate of strong sphaleron
transitions is comparable to or even larger than the rate of
chirality-flip transitions through the Yukawa
coupling of the top quark. Hence, these
processes must be taken into account. This decreases the
number of conserved currents by one\footnote{Note that
influence of strong sphalerons on a different mechanism for
electroweak baryogenesis has been
discussed also in ref. \cite{mohapatra}.}.\\

(iii) Anomalous fermion number non-conservation must be taken into
account \cite{krs}. It decreases the number of conserved currents by
one.\\

Therefore, a complete set of conserved anomaly-free fermionic currents
can be
represented as
\ba
\label{first}&&\bar{Q_i}\gamma_{\mu}Q_i + R - L_L,~ i = 1,2,\\ &&
\bar{Q_3}\gamma_{\mu}Q_3 + \bar{U_3}\gamma_{\mu}U_3 + R/2 - L_L,\\&&
\bar{L_i}\gamma_{\mu}L_i - L_L/3,i = 1,2,\\&&
\bar{E_i}\gamma_{\mu}E_i, i = 1,2,3,\\&&
\bar{D_i}\gamma_{\mu}D_i -R/2, i=1,2,3,\\&&
\bar{U_1}\gamma_{\mu}U_1 - \bar{U_2}\gamma_{\mu}U_2,
\label{charges}
\ea
where
\eq
L_L = \sum_{i=1}^3 \bar{L_i}\gamma_{\mu}L_i,~
R = \sum_{i=1}^2 \bar{U_i}\gamma_{\mu}U_i.
\en
In addition to the quantum numbers associated with these currents,
there exists a conserved operator containing scalar fields. This is the
familiar hypercharge
\eq
Y = Y_F + Y_s,
\en
where
\eq
Y_s = - i [\phi_1^{\dagger}{\cal{D}}_0\phi_1  -
({\cal{D}}_0\phi_1)^{\dagger}\phi_1] - i [1 \rightarrow 2]
\en
is the scalar field contribution. Finally, one should add the
third component of the
weak isospin operator $T_3$. As in \cite{ckn1,ckn3}, we assume that
the classical motion of
the scalar field does not introduce any non-zero hypercharge or $T_3$
density. Therefore the hypercharge density is equal to zero
for a uniform spinodial decomposition phase transition.

Now one can check that, in the massless quark
approximation, eqs. (\ref{sum}, \ref{density}, \ref{equations},
\ref{first} - \ref{charges}) imply that the equilibrium value of
the
baryonic charge is zero, $\langle B \rangle = 0$. In order to get this
result, we used, following ref. \cite{khlsh}, the particle
spectrum of the {\em unbroken} phase rather than the broken one.

The reader may wonder why we did not use the electric charge operator
and the physical particle spectrum of the broken
phase instead of
$Y$ and $T_3$ \footnote{Actually, the zero result is reproduced also
with $Q$
instead of $T_3$ and $Y$ and the particle spectrum of the broken
phase.}.
The reason is that the use of the {\em physical} spectrum in such a
calculation would correspond to a computation of the free
energy of the system in the broken phase in the one-loop approximation
in
a  {\em unitary gauge}. It is known \cite{dolan,weinberg,arnold}, that
perturbation theory in a unitary gauge is a very delicate thing
({\em e.g.}, it
does not converge \cite{arnold}). In particular, it gives an incorrect
result for the critical temperature of the phase transitions in gauge
theories.

It is clear that for sufficiently small vacuum expectation values of
the
Higgs field, all effects associated with spontaneous symmetry breaking
are
just the mass corrections which can be neglected in the
high-temperature
approximation in any {\em renormalizable gauge}. In the worst case, the
expansion parameter may be $(v/T)^2$. In our case, the vacuum
expectation
value of the Higgs field $v$ is bounded from above by the requirement
that the sphaleron processes are sufficiently fast, so that $v < g_W
T$.
Therefore, the use of the particle spectrum of the unbroken phase is
perfectly justified.

The zero result is also reproduced when
fermionic number non-conservation is out of thermal equilibrium. Then
one can write an equation for the baryonic number evolution
\cite{khlsh} (see also refs. \cite{ckn1,ckn2,ckn3})
\eq
\frac{\partial B}{\partial t} = - N^2
\frac{\Gamma_{sph}}{T} \frac{\partial F}{\partial B} =
-9\frac{\Gamma_{sph}}{T}\mu_B,
\label{dilution}
\en
where $F(B)$ is the free energy of the system characterized by zero
values
of all conserved charges but non-zero baryonic density $B$,
$\Gamma_{sph}$
is the rate of sphaleron transitions, and $N=3$ is the number of
fermionic
generations.  In our case,
\eq
\frac{\partial F}{\partial B} = \mu_B \sim B,
\en
so that $B$ stays zero all the time.

We stress that the inclusion of
strong sphalerons was essential for this result. Strong
sphalerons have the physical effect of mantaining
the same chemical potential for left- and right-handed
baryonic numbers. If instead one neglects these processes,
then the system of charges (\ref{charges}) has to be supplied with an
extra charge, say $R$, while the density matrix will contain an
additional chemical potential. One can check that if one
chooses this set of conserved charges, the result is indeed non-zero.
Which solution has to be used? The answer depends on the ratio of the
typical time scales. If $\tau_0 \ll \tau_{strong}$, then strong
sphalerons are irrelevant, and the asymmetry indeed does not contain
Yukawa coupling suppression provided that $\tau_Y$ is smaller than the
typical
time scale $\tau_0$ of the Higgs phase change in the
spontaneous baryogenesis mechanism.
If, on the contrary, $\tau_0 > \tau_{strong}$ or $\tau_0 \sim
\tau_{strong}$, then the asymmetry vanishes in the massless
approximation.

In the original calculation of
Cohen, Kaplan and Nelson \cite{ckn1,ckn2} a non-zero result for
the baryonic asymmetry was obtained, since strong sphaleron
transitions have been neglected and only
$B-L$, $Q$, and  $B_3 - \frac{1}{2}(B_1+ B_2)$
have been included as conserved charges. However, in presence
of strong sphalerons transitions, their solution corresponds
to non-zero averages for some of the approximately
{\em conserved}
charges defined in eqs. (\ref{first} - \ref{charges}).

Is the conclusion that $\langle B \rangle = 0$ fatal for the
spontaneous baryogenesis mechanism? In fact, this is
not the case when quark mass corrections are included.
Mass corrections can be taken into account in perfect
analogy
with the case of the leptonic asymmetries discussed in refs.
\cite{krs1,khlsh}
(for a later discussion see ref. \cite{dreiner}). To make the
computation simpler (and to make clearer why we previously obtained
zero),
we observe that in fact the number of
independent chemical potentials (14) can be reduced to a smaller
number (6) with the use of flavour symmetry. Namely, our lagrangian
has quite a large global symmetry group $SU(2)_{Q}$ x $SU(3)_{D}$ x
$SU(2)_{U}$ x
$SU(3)_{L}$ x $SU(3)_{E}$, where the group labels denote the
fields on which
the corresponding group is acting. The conserved charges invariant
under
this symmetry are
\ba
 \label{first1}&& A_1 =
\sum_{i=1}^2 \bar{Q_i}\gamma_{\mu}Q_i +2 R -2 L_L,\\&&
A_2 =
\bar{Q_3}\gamma_{\mu}Q_3 + \bar{U_3}\gamma_{\mu}U_3 + R/2 - L_L,\\&&
A_3 =
\sum_{i=1}^3 \bar{D_i}\gamma_{\mu}D_i -3R/2,\\&&
A_4 =
\sum_{i=1}^3\bar{E_i}\gamma_{\mu}E_i,
\label{last1}
\ea
hypercharge $Y$ and weak isospin $T_3$. We denote the corresponding
chemical potentials by
$\mu_i,~i$ = 1, ... , 4, $\mu_Y$ and $\mu_T$. They are to be found
from
the equations
\eq
\langle A_i \rangle = \langle Y \rangle = \langle T_3 \rangle = 0.
\en
Let us first solve this system neglecting the mass of the top quark.
In order to show that
the baryonic density is zero, one can consider only
the total baryonic density and the left-handed
leptonic density. The total baryonic density is
\eq
\langle B \rangle = 4[(\dot{\theta}+\mu_Y) + 2 \mu_1 +
\mu_2]\frac{T^2}{6},
\en
while the total left leptonic density is
\eq
\langle L_L \rangle = -6 [(\dot{\theta}+\mu_Y) + 2 \mu_1 +
\mu_2]\frac{T^2}{6}.
\en
Then from $\langle B-L_L \rangle = 0$, it follows $\langle B
\rangle=0$.
For future reference, we give here the complete solution of eq. (30)
in the massless approximation:
\ba
&&\mu_1 = - \frac{4}{21}(\dot{\theta}+\mu_Y),~
\mu_2 = - \frac{13}{21}(\dot{\theta}+\mu_Y),~
\mu_3 =  \frac{11}{21}(\dot{\theta}+\mu_Y),\\&&
\mu_4 =  2 (\dot{\theta}+\mu_Y),~\mu_T = 0,~
(\dot{\theta}+\mu_Y) = \dot{\theta}\frac{14 n_s}{9+14 n_s },
\ea
where $n_s$ is the number of scalar doublets.

The exact proportionality between baryonic number and left-handed
leptonic
number is, however,  an artifact of the massless approximation.
Using eq. (\ref{quark}), the baryonic density becomes
\eq
\langle B \rangle = - \frac{3m^2}{20 \pi^2 T^2} \rho_Y = -
\frac{9n_s}{10(9+14n_s )\pi^2}
m^2\dot{\theta},
\label{final}
\en
where $\rho_Y$ is the fermionic hypercharge density. This is our final
result for the equilibrium density of the baryonic
number in the background of a scalar field with slowly changing phase.

If strong sphalerons were not in equilibrium, we should add a new
approximately conserved charge, for instance $A_5\equiv R$, to our
previously defined set of charges, see eqs.(26)--(29). Following
the same procedure used to obtain eq.(35), we would find:
\eq
\langle B \rangle =\frac{n_s}{6+11n_s}T^2\dot{\theta}.
\en
Therefore the presence of strong sphalerons leads to a suppression
factor $\frac{126}{185}(\frac{m}{\pi T})^2$, where we have taken
$n_s=2$; this corresponds numerically to a suppression of about
$3\cdot 10^{-2}$ for $f_t \sim 1$, $v(T) \sim g_W T$.

The approximation that anomalous baryon number violation is in
equilibrium is unlikely to be correct. Thus, following ref.
\cite{ckn1},
we consider the evolution equation for the baryon asymmetry density,
eq. (24). We
can compute $\mu_B$
using the same procedure followed above, adding $B$ to the set of
conserved charges and imposing, besides eq. (30), the further
constraint $\langle B \rangle =0$. We find, if strong sphalerons
are in equilibrium,
\eq
\mu_B=
\frac{9n_s}{4(9+14n_s )}\frac{m^2}{\pi^2T^2}
\dot{\theta},
\en
and, if strong sphalerons are not in equilibrium,
\eq
\mu_B=-
\frac{2n_s}{3(1+2n_s )}
\dot{\theta}.
\en
The mass suppression present when strong sphalerons are in
equilibrium amounts to a factor $\frac{135}{296}(\frac{m}{\pi T})^2
\sim 2\cdot 10^{-2}$.

We can now consider the case of baryogenesis with the charge transport
mechanism.
As shown in ref. \cite{ckn2}, the CP-violating interactions of the
fermions with the moving thin bubble wall lead to a non-vanishing
flux of particles carrying some quantum number $X$. If, as in the
original work of ref. \cite{ckn2}, we identify $X$ with the
hypercharge,
then it is easy to realize that $\langle B\rangle =0$, when strong
sphalerons are in equilibrium and all fermions are massless. The
calculation is analogous to the one for spontaneous baryogenesis, with
the only difference that, instead of the source $\dot{\theta}$ we
have to consider a non-vanishing $Y$ density. However, as shown by
Khlebnikov \cite{khlebnikov}, the Debye screening prevents the
transport
of the gauged charge $Y$ over distances larger than about $2/T$.

Therefore the recipe proposed in ref.
\cite{ckn4} is to identify as the transported charges $X$ only global
charges which are orthogonal to gauge charges. We then define
\eq
B^{\prime}=B+\alpha_B Y,~~~~~~
A_i^{\prime}=A_i+\alpha_i Y,~~~~~~i=1,...,4,
\label{char}
\en
with $\alpha_B$ and $\alpha_i$ chosen such that $B^\prime$ and
$A_i^\prime$ are orthogonal to $Y$, {\em i.e.}, in a massless quark
approximation:
\eq
-\alpha_B=-\frac{1}{8}\alpha_1=-\frac{1}{4}\alpha_2=
\frac{2}{9}\alpha_3=\frac{2}{3}\alpha_4=\frac{1}{10+n_s}.
\en

By introducing chemical potentials for the charges (26)--(29),
$Y$, and $B$, we
can write the asymmetry density for each species of particles as:
\eq
\rho=\frac{T^2}{6}\left( q_Y \mu_Y + q_B\mu_B
+\sum_i q_{A_i} \mu_i\right) .
\en
The chemical potentials $\mu_Y$, $\mu_B$, and $\mu_i$ are then
computed from the equations
\eq
\langle Y \rangle =0,~~~\frac{\langle B \rangle}{\alpha_B}
=\frac{\langle A_i \rangle}{\alpha_i} = n_x \neq 0,
\label{sol}
\en
which correspond to a situation in which the gauged charge $Y$ is
screened, but the $Y$ components of the global charges can carry
a non-vanishing density $n_x$. The solution of eq. (\ref{sol}) yields
$\mu_B =0$ and therefore anomalous baryon number violating
interactions are unable to generate any baryon asymmetry. One
can also check that $\mu_B \neq 0$, if strong sphalerons are not
in equilibrium.

The cancellation in the case of charge transport baryogenesis may
seem more severe than in the case of spontaneous baryogenesis since
the relevant processes occur
in the {\em unbroken phase}, where the Higgs field condensate is
absent.
There are two types of
corrections which save the situation. Just as in the leptonic case
considered in \cite{khlsh}, Yukawa and gauge radiative corrections are
important.
The equation for the asymmetry number density of each chiral fermion
in the unbroken phase, including radiative corrections, can still be
written as in eq.(12), with the replacement:
\eq
m^2\to \left[ C_sg_s^2+(C_W+\frac{Y^2}{4}\sin^2\theta_W)g_W^2+
\frac{f^2}{2}\right] \frac{T^2}{4},
\label{masses}
\en
where $C_s$ is $\frac{4}{3}$ for quarks and $0$ for leptons,
$C_W$ is $\frac{3}{4}$ for weak doublets and $0$ for singlets,
$Y$ is the hypercharge quantum number, and $f$ is the Yukawa
coupling, taken to be non-vanishing only for the top quark.
The solution of eq. (\ref{sol}) now gives:
\eq
\mu_B  = \frac{9}{64 \pi^2(9 + 14 n_s)}(3f_t^2 + 2 g_W^2
\sin^2\theta_W)
\frac{n_x}{T^2},
\label{pif}
\en
which, as noticed above, vanishes in the limit of negligible
radiative corrections. The result of eq. (\ref{sol}) would
correspond to a reduction of
the  baryonic asymmetry in comparison with
\cite{ckn2,ckn3} by the factor $\sim \frac{27(1+2n_s)}{128
\pi^2(9+14n_s)}f_t^2 \sim 3\cdot 10^{-3}$, if other effects are absent.
These
more important effects
are associated with  electroweak corrections which break the
degeneracy between left and right top quarks in the unbroken phase
resulting in a net baryonic current $J_B$  (see refs. \cite{msm,far}) related
to
the hypercharge current $J_Y$ through
\eq
J_B = \frac{2}{3}\frac{p_L -  p_R}{p_L +  p_R}J_Y,
\en
where $p_L$ and $p_R$ are the momenta of incident left and right quarks
corresponding to the same energy. Now, $p_L$ and $p_R$ are different since left
quarks interact with SU(2) gauge fields while right quarks do not. With the use
of eq. (\ref{masses}) we get  a larger baryonic chemical potential,
\eq
\mu_B \simeq \frac{\alpha_W \pi }{16 }\frac{n_x}{\bar{p}^2},
\en
where ${\bar{p}}>m_t$ is a typical transverse momentum  of the quarks
contributing to charge transport.
So the final estimate of the suppression of the asymmetry in this case
is a factor of about $10^{-2}$. Therefore, the  charge transport
mechanism
(as well as the ``topological'' mechanism \cite{mstv,gst}) can still
explain the baryonic asymmetry since, according to the estimate
of ref. \cite{ckn2},
$n_B/s$ can be as large as $10^{-7}$. With strong sphalerons taken
into account, this number should be converted to $10^{-9}$,
which
is still larger than observation.

To conclude, we have shown that the baryonic asymmetry in the two Higgs
doublet model produced by any of the mechanisms considered in
\cite{ckn1,ckn2,ckn3} contains a parametric
suppression associated with the top Yukawa coupling or electroweak
coupling due to the existence of
strong sphalerons. Strong sphalerons  have not been
taken into account in a number of papers on electroweak baryogenesis
and may change their conclusions.

One of us (M.S.) is indebted to the High Energy Physics Group at
Rutgers
University, where part of this this work has been done, for kind
hospitality.
We thank G. Farrar, D. Kaplan, S. Khlebnikov, A. Nelson  and N. Turok
for useful discussions.

\end{document}